\documentclass[pdflatex,sn-nature]{sn-jnl}

\usepackage{graphicx}%
\usepackage{multirow}%
\usepackage{amsmath,amssymb,amsfonts}%
\usepackage{amsthm}%
\usepackage{mathrsfs}%
\usepackage[title]{appendix}%
\usepackage{xcolor}%
\usepackage{textcomp}%
\usepackage{manyfoot}%
\usepackage{booktabs}%
\usepackage{lineno}
\usepackage{subcaption}
\usepackage{float}
\usepackage{tabularx}
\usepackage{caption}
\begin{document}

\title[Article Title]{Composite spectrum of Little Red Dot from a standard inner disk and an unstable outer disk}

\author[1]{\fnm{Chenxuan} \sur{Zhang}}

\author*[1]{\fnm{Qingwen} \sur{Wu}} \email{qwwu@hust.edu.cn}

\author[2]{\fnm{Xiao} \sur{Fan}}

\author[3,4]{\fnm{Luis C.} \sur{Ho}}

\author[1,5]{\fnm{Jiancheng} \sur{Wu}}

\author[1]{\fnm{Huanian} \sur{Zhang}}

\author[3]{\fnm{Bing} \sur{Lyu}}

\author[5]{\fnm{Xinwu} \sur{Cao}}

\author[6,7,8]{\fnm{Jianmin} \sur{Wang}}

\affil[1]{\orgdiv{Department of Astronomy, School of Physics}, \orgname{Huazhong University of Science and Technology}, \orgaddress{\street{Luoyu Road}, \city{Wuhan}, \postcode{430074}, \state{Hubei}, \country{China}}}

\affil[2]{\orgdiv{School of Physics and Technology}, \orgname{Wuhan University}, \orgaddress{\street{Luoyu Road}, \city{Wuhan}, \postcode{430072}, \state{Hubei}, \country{China}}}

\affil[3]{\orgdiv{Kavli Institute for Astronomy and Astrophysics}, \orgname{Peking University}, \orgaddress{\street{Yiheyuan Road}, \city{Beijing}, \postcode{100871}, \state{Beijing}, \country{China}}}

\affil[4]{\orgdiv{Department of Astronomy, School of Physics}, \orgname{Peking University}, \orgaddress{\street{Yiheyuan Road}, \city{Beijing}, \postcode{100871}, \state{Beijing}, \country{China}}}

\affil[5]{\orgdiv{Institute for Astronomy, School of Physics}, \orgname{Zhejiang University}, \orgaddress{\street{Yuhangtang Road}, \city{Hangzhou}, \postcode{310027}, \state{China}, \country{China}}}

\affil[6]{\orgdiv{Key Laboratory for Particle Astrophysics, Institute of High Energy Physics}, \orgname{Chinese Academy of Sciences}, \orgaddress{\street{Yuquan Road}, \city{Beijing}, \postcode{100049}, \state{Beijing}, \country{China}}}

\affil[7]{\orgdiv{National Astronomical Observatories of China}, \orgname{Chinese Academy of Sciences}, \orgaddress{\street{Datun Road}, \city{Beijing}, \postcode{100101}, \state{Beijing}, \country{China}}}

\affil[8]{\orgdiv{School of Astronomy and Space Science}, \orgname{University of Chinese Academy of Sciences}, \orgaddress{\street{Yuquan Road}, \city{Beijing}, \postcode{100049}, \state{Beijing}, \country{China}}}

\abstract{James Webb Space Telescope (JWST) has revealed a new class of high-redshift, very red, compact broad-line sources, termed as “little red dots” (LRDs). The physical mechanism driving these properties remains elusive. We construct spectral energy distributions (SEDs) with spectroscopic redshift for 28 LRDs and find they exhibit V-shaped SEDs with a common break frequency of $\nu_{\rm b}\simeq10^{14.96\pm0.06}$ Hz. We propose that the unique SEDs can be well explained by the combination of an inner standard disk and an outer gravitationally unstable accretion disk with Toomre parameter $Q\sim1$, where the outer disk has a temperature of $\sim2000-4000 K$ and mainly radiates in near-infrared to optical wavebands. The composite spectrum from this model naturally explains the V-shaped continuum and reproduces intrinsically luminous infrared–optical emission without requiring extreme dust extinction or unusual stellar populations. Even considering possible dense gas around the disk to account for pronounced Balmer breaks in some LRDs, the intrinsic optical–UV emission is only suppressed by factors of $\lesssim2-3$, which suggests that most LRDs are sub-Eddington and intrinsically weak. These results provide new insights into early-phase black hole growth and galaxy evolution.}

\maketitle
\section*{Main Text}\label{main}
James Webb Space Telescope (JWST) is opening up a new era in the study of the early Universe, thanks to its remarkable sensitivity and angular resolution at infrared wavelength \cite{2023PASP..135f8001GardnerBrammer}. One of the most surprising results from JWST is the discovery of numerous faint, optically red and extremely compact sources ($R_\mathrm{e} \approx 100$ pc) \cite{2023ApJ...942L..17Onoue, 2025ApJ...986..126Kocevski, 2025ApJ...983...60Chen}. These sources are referred to as “Little Red Dots” \cite{2023Natur.616..266Labbe, 2024ApJ...964...39Greene, 2024ApJ...963..129Matthee, 2025ApJ...986..126Kocevski}, which are widely considered as candidates for high-redshift active galactic nuclei (AGNs) based on their broad emission lines, and can play an important role in understanding the evolution of galaxies and supermassive black holes at cosmic dawn \cite{2024ApJ...964...39Greene, 2024ApJ...963..129Matthee, 2023ApJ...954L...4Kocevski, 2025arXiv250316595Rusakov, 2025ApJ...990..160Wang, 2025ApJ...986..165Taylor}. The LRDs exhibit spectral energy distributions (SEDs) characterized by a ``V-shaped" continuum with a luminous red component and a faint blue component in the rest-frame spectrum, which is notably different from that of local AGNs \cite{2023Natur.616..266Labbe, 2024ApJ...963..129Matthee, 2025ApJ...986..126Kocevski}. Furthermore, the LRDs are less variable and are X-ray weaker compared to classical AGNs \cite{2024ApJ...974L..26Yue, 2024ApJ...969L..18Ananna,2025ApJ...985..119Zhang}. Interpreting the LRD SEDs remains difficult, which prevents us from further understanding the possible SMBH growth or galaxy evolution in high redshift \cite{2023ARA&A..61..373Fan, 2024ApJS..275...19Ye}. The common AGN-only or galaxy-only models cannot adequately describe the two distinct spectral components in V-shaped SEDs. For the AGN-only scenario, the red/optical component is dust-reddened AGNs due to the dust in the torus and/or interstellar medium \citep[e.g.,][]{2024A&A...691A..52Killi, 2023ApJ...954L...4Kocevski}. It has also been proposed that the LRDs may be powered by the pure dusty starburst galaxies \cite{2025ApJ...984..121Wang, 2024ApJ...968....4Perez-Gonzale, 2024ApJ...968...34Williams, 2025ApJ...981..191Ma}. Alternatively, some studies also consider the co-existence of AGNs and galaxies: either a galaxy dominates the rest-frame UV SED while a reddened AGN contributes in the rest-frame infrared to optical range; or the AGN contributes significantly to the rest-frame UV while a dusty stellar population contributes in the rest-frame UV SED \cite{2024A&A...691A..52Killi, 2025ApJ...984..121Wang, 2025ApJ...981..191Ma}. A fraction of LRDs exhibit a clear Balmer break in their spectra, which indicates either an old stellar population \cite{2024ApJ...977L..13Baggen, 2024ApJ...969L..13Wang} or extremely dense gas surrounding an AGN \cite{2025ApJ...980L..27Inayoshi, 2025A&A...701A.168deGraaff, 2025MNRAS.tmp.1770Ji}. Most of these models predict significant dust emission in LRDs to explain their distinctive V-shaped SEDs. However, observations reveal no detectable mid- to far-infrared emission, suggesting a lack of both hot and cold dust components \cite{2025ApJ...991L..10Setton, 2025A&A...700A.231Xiao,2024ApJ...975L...4Casey}, which contradicts the attenuated luminosities as anticipated in these models.

\begin{figure}
    \centering
    \includegraphics[width=\textwidth]{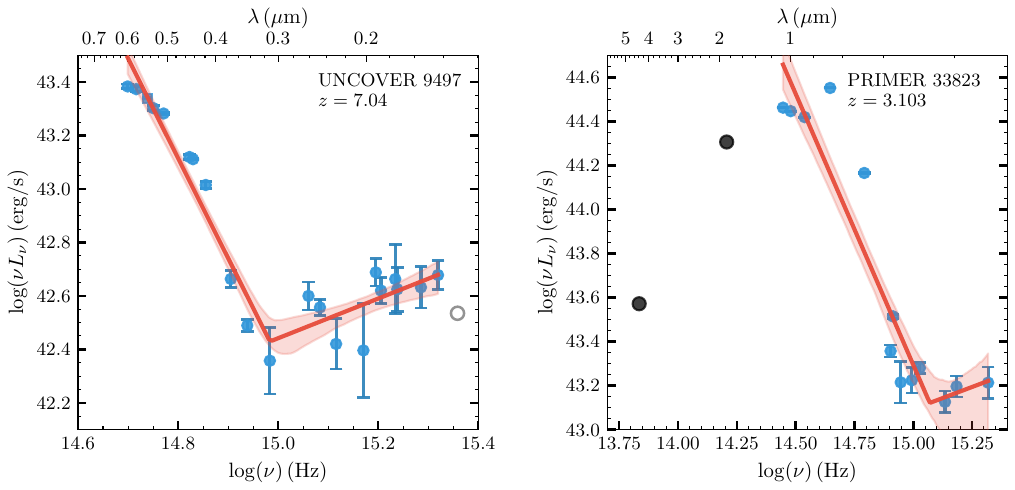}
    \caption{\textbf{Broken-power-law fits for two typical LRDs.} The blue points with 1 sigma error bars indicate the photometric data used for fitting. The black points are MIRI measurements excluded from the fit, and the grey-outlined white point in the left panel marks the Ly$\alpha$-blueward flux that is also masked. The red line with the shaded region shows the best fits with $1\sigma$ uncertainty.}
    \label{figure_1}
\end{figure}

\begin{figure}
    \centering
    \includegraphics[width=1.0\textwidth]{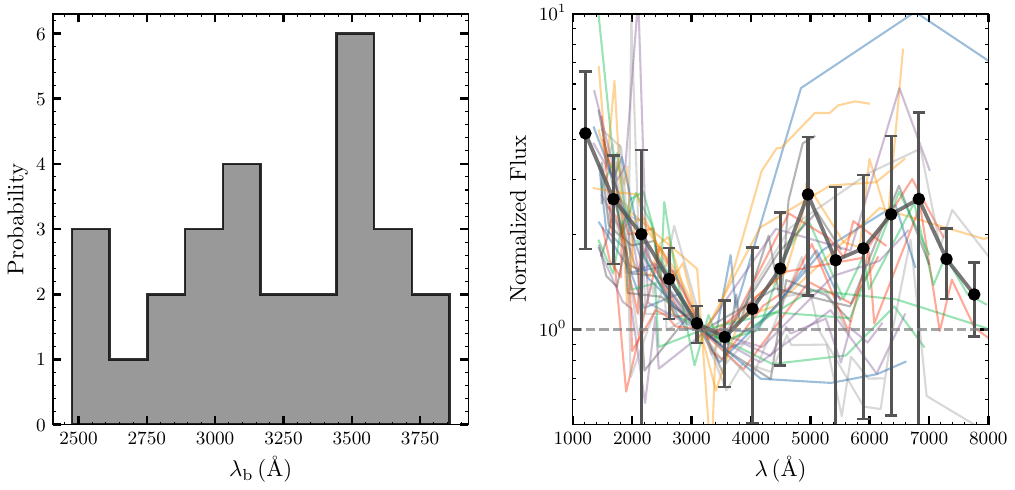} 
    \caption{\textbf{Distribution of Break wavelengths and normalized SEDs of LRDs.}The left panel shows the histogram of the break wavelength in V-shaped LRD SEDs. The right panel shows the normalized SEDs at $\lambda_{\rm b} = 3287\,$\AA, corresponding to the break frequency $\nu_{\rm b}=10^{14.96}$ Hz, where the binned value with standard deviation is shown as black dots.}
    \label{figure_2}
\end{figure}

In this paper, we build the SEDs based on the photometric and spectroscopic data for a sample of 28 LRDs, all exhibiting broad emission lines. We firstly fit the V-shaped SED in the rest frame with a simple broken-power-law model (see Methods for the fittings). We present the fittings for two LRDs, UNCOVER 9497 and PRIMER 33823, as examples in Figure \ref{figure_1}, which illustrate the characteristic V-shaped SEDs and bumped infrared-to-optical SEDs, respectively. In the linear fitting, we exclude photometric data points with frequencies higher than the Lyman limit, Signal-to-Noise Ratio (SNR) less than 1, or data significantly affected by emission lines. Additionally, the long-wavelength MIRI data of PRIMER 33823 are disregarded in the linear fitting due to their evident deviation from the V-shaped spectrum, which will be addressed in a more physically motivated model. Interestingly, we find that the break frequencies in the V-shaped SEDs fall within a narrow range, from $\sim 10^{14.8}$ to $10^{\sim 15.1}$ Hz (or the inflection wavelength from 2600-3800 \AA). We present the distribution of the inflection wavelength in the left panel of Figure \ref{figure_2}, where $\lambda_\mathrm{b} = 3287^{+487}_{-424}$ \AA \ or $\nu_\mathrm{b} = 10^{14.96 \pm 0.06}$ Hz. These fitting results suggest LRDs exhibit a similar break frequency in the rest-frame V-shaped SED. We present the SEDs for 28 LRDs in the right panel of Figure \ref{figure_2}, which are normalized at the inflection wavelength $\lambda_\mathrm{b}$.

\begin{figure}
    \centering
    \includegraphics[width=0.6\linewidth]{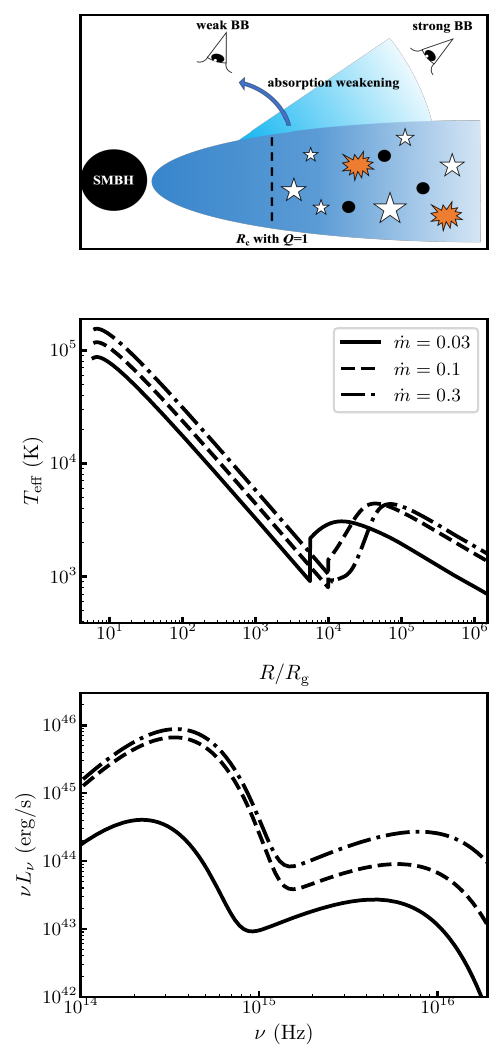}
    \caption{\textbf{Illustration of the two-component accretion-disk model.} The top panel shows a schematic diagram of the inner standard disk and the outer gravitationally unstable disk with disk winds. The distribution of the disk temperature and the corresponding SED are shown in the middle and bottom panels for $M_{\rm BH}=10^{7.5}M_{\odot}$ and accretion rates of $\dot{m}=0.03, 0.1, 0.3$. This model predicts V-shaped SEDs, arising from the inner standard thin disk and the outer marginally unstable disk.}
    \label{figure_3}
\end{figure}

We explore the physical mechanism underlying this special feature of SEDs. The standard accretion disk is believed to power luminous AGNs \cite{1973A&A....24..337Shakura}. It is well known that the outer part of the standard disk, beyond a critical radius is prone to suffer gravitational instability if Toomre parameter $Q=c_{\rm s}\Omega/\pi G \Sigma < 1$, where $c_{\rm s}$, $\Omega$, $G$ and $\Sigma$ are sound speed, angular velocity, gravitational constant and disk surface density respectively \cite{1964ApJ...139.1217Toomre}. In AGNs, the gravitational instability of the accretion disk provides a mechanism for forming nuclear stars and enriching the metals in the central part of galaxies as indicated by the broad emission lines \citep[e.g.,][]{1987Natur.329..810Shlosman, 2003MNRAS.339..937Goodman, 2023ApJ...954...84Wang, 2023ApJ...944..159Fan}. The feedback from the stellar process (e.g., supernova, stellar wind) and turbulent heating can balance radiative cooling of the accretion disk, allowing the outer unstable disk to eventually self-regulate at $Q\sim1$ \citep[e.g.,][]{2003MNRAS.341..501Sirko, 2022ApJ...928..191Gilbaum, 2023ApJ...948..120Chen}, where the additional heating will increase the disk temperature in the outer disk and mainly radiate in the near infrared to optical wavebands \cite{2003MNRAS.341..501Sirko}. The structure of the accretion disk and the radiation spectrum can be derived from the SMBH mass, $M_{\rm BH}$, accretion rate,  $\dot{M}$, viscosity parameter, $\alpha$, and outer radius of the accretion disk, $R_\mathrm{out}$. In this work, the viscosity parameter $\alpha=0.1$ is adopted. In Figure \ref{figure_3}, we present a cartoon picture (top panel), disk temperature (middle panel), and emitted spectrum (bottom panel) of this disk for typical AGN parameters. The disk temperature of the outer marginally unstable disk is around 2000-4000 K in the range of $10^{4-6}R_{\rm g}$ ($R_{\rm g}=GM_{\rm BH}/c^2$ is gravitational radius, and $c$ is light speed), which is much hotter than the typical temperature of dust torus. The blackbody emission of the outer disk with this temperature mainly radiates in the near infrared to optical waveband, which can explain the red excess in the spectra of the LRDs. This model predicts a V-shaped SED, in which the near-infrared to optical and optical to UV emission originates from the outer marginally unstable disk and the inner standard accretion disk, respectively. Both the pronounced Balmer breaks and absorption features superimposed on their Balmer emission lines suggest the possible presence of dense absorbing gas surrounding some LRDs \cite{2024ApJ...973L..49Inayoshi, 2024ApJ...963..129Matthee, 2024Natur.636..594Juodzbalis, 2025ApJS..277....4DEugenio, 2025MNRAS.tmp.1770Ji}. Following \cite{2024ApJ...973L..49Inayoshi, 2025MNRAS.tmp.1770Ji}, we adopt \textsc{Cloudy} \cite{2023RMxAA..59..327Chatzikos} to model the Balmer break caused by the transmission of the intrinsic AGN continuum through a dense gas (see Methods).

\begin{figure}
    \centering
    \includegraphics[width=1.0\textwidth]{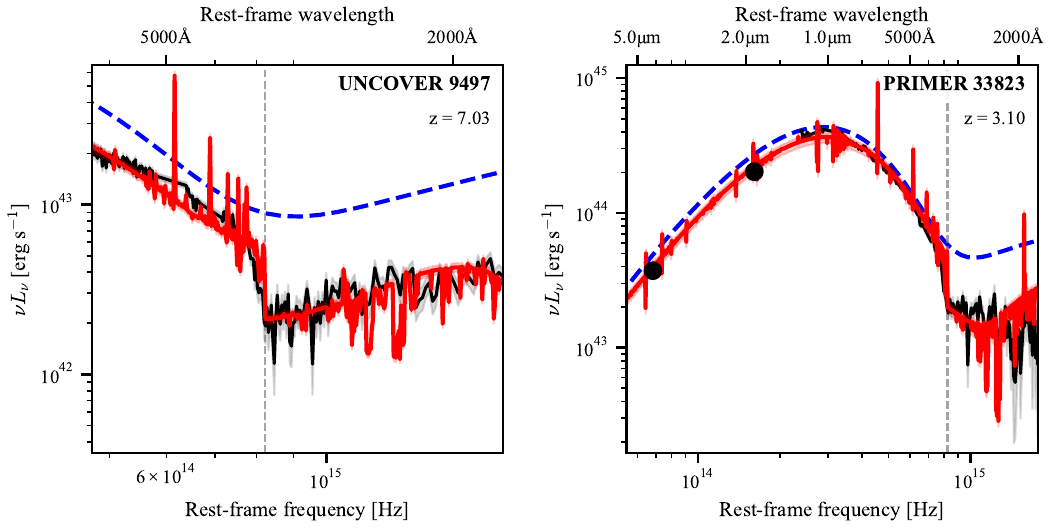}
    \caption{\textbf{MCMC fits of two typical LRD SEDs.} The black solid lines show the observed spectra, the red solid lines indicate the best-fit absorbed SEDs, and the blue dashed lines represent the intrinsic SEDs before absorption. The grey-shaded regions denote the $1\sigma$ uncertainty bands derived from the MCMC posterior distributions. The vertical dotted line marks the Balmer limit at 3645\,\AA.}
    \label{figure_4}
\end{figure}

\begin{figure}
    \centering
    \includegraphics[width=1.0\textwidth]{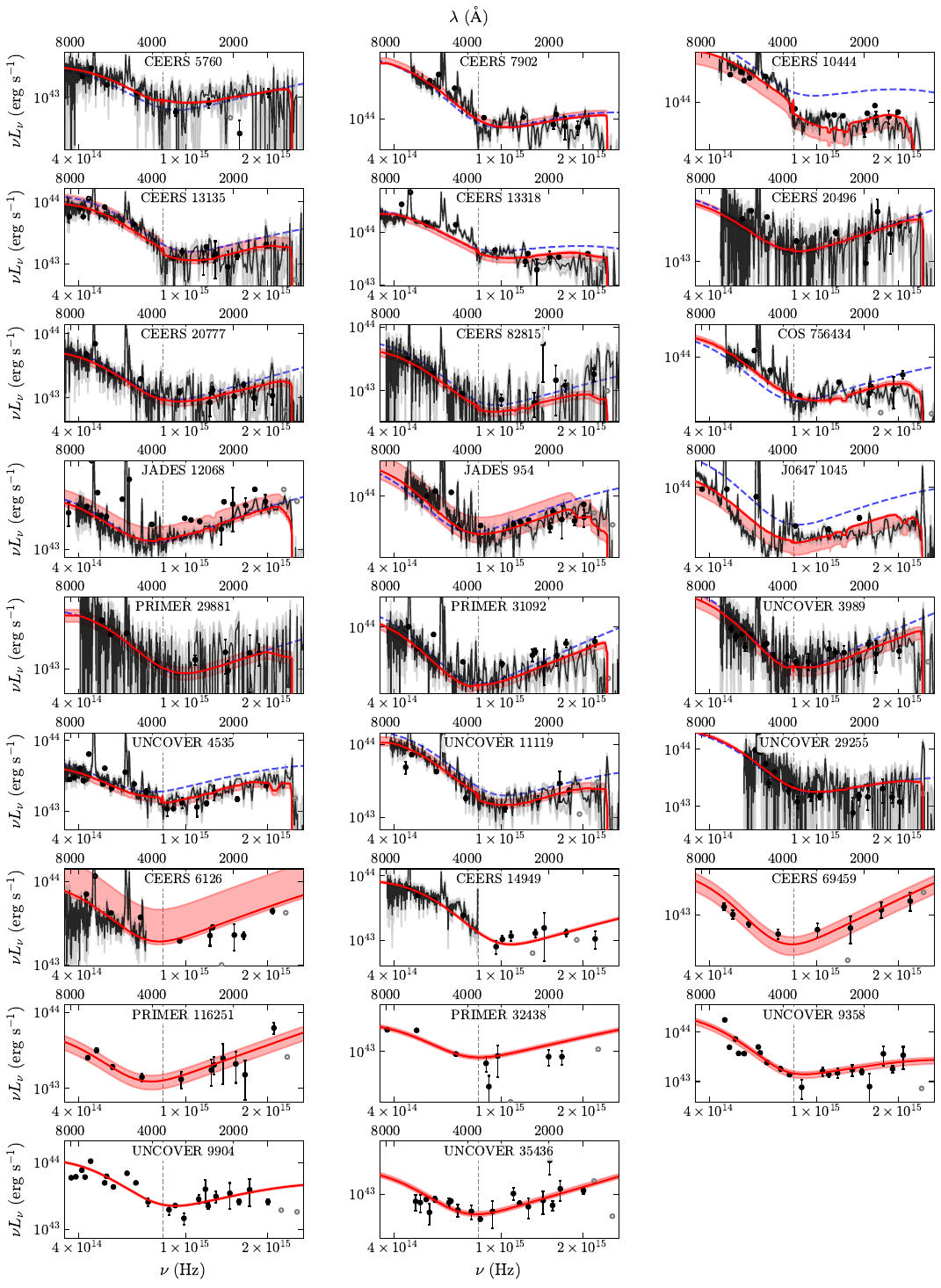} 
    \caption{\textbf{SED modeling of V-shaped LRDs.} Black points denote the measured fluxes, with error bars indicating the 1$\sigma$ photometric or spectroscopic uncertainties. Those with SNR $<1$ or affected by emission lineare shown as open grey circles. The red solid curve shows the best-fitting absorbed SED obtained from the MCMC analysis, and the red shaded region represents the 1$\sigma$ uncertainty band derived from the posterior distribution. For sources modeled with four free parameters, the intrinsic (unabsorbed) model SED is plotted as a blue dashed line.}
    \label{figure_5}
\end{figure}

We then fit the LRD SEDs using the above physical model employing the Markov chain Monte Carlo (MCMC) method implemented in emcee \cite{2013PASP..125..306Foreman}. The adopted SMBH masses in our sample are estimated from broad emission lines, and this method is roughly consistent with that directly estimated from dynamical mass measurement in a strongly lensed LRD \cite{2025arXiv250821748Juodzbalis}.  We treat the other four parameters, accretion rate $\dot{M}$, outer radius of disk $R_\mathrm{out}$, and gas density $n_{\rm H}$, column density $N_{\rm H}$ of dense absorber as free variables for 20 LRDs with well-sampled spectroscopic data and well-constrained Balmer breaks. Gas absorption is neglected in the other 8 LRDs with poor spectroscopic data and poorly constrained Balmer breaks (see Method for more details). In Figure \ref{figure_4}, we present an example of the modeling results for two LRDs, UNCOVER 9497 and PRIMER 33823, with a prominent Balmer break, while the fitting for the other 26 sources is presented in Methods. The best-fit parameters for UNCOVER 9497 and PRIMER 33823, including the accretion rate $\dot{M}$, outer disk radius $R_\mathrm{out}$, hydrogen number density $n_{\rm H}$, and column density $N_{\rm H}$, are summarized in Table 1, where the posterior distributions of parameters are shown in Supplementary Figure \ref{supfig:supp_figure_3}. For UNCOVER 9497, the Balmer break strength is $BBs=3.47$, indicating strong absorption in the optical-UV band, while PRIMER 33823 shows $BBs=2.51$, with moderate absorption. Overall, the intrinsic optical-UV emission is suppressed by factors of 1.5–3.2 at 2500–5100,\AA\ in sources with strong Balmer breaks. The full set of parameters for all 20 well-sampled LRDs is listed in Table 1, showing hydrogen number densities $n_{\rm H} \sim 10^{7-9}\rm,cm^{-3}$ and column densities $N_{\rm H} \sim 10^{22-25}\rm,cm^{-2}$ for the dense absorber. Absorption is generally negligible for sources with weak Balmer breaks (see Figure \ref{figure_5}). The inferred hydrogen densities ($n_{\rm H}$) in our model are roughly an order of magnitude lower than the extreme values reported in \cite{2025MNRAS.tmp.1770Ji, 2025arXiv250316596Naidu}. This reduction is expected due to the adoption of an intrinsic V-shaped SED, which naturally accounts for part of the continuum suppression without requiring extremely high gas densities. The current model occasionally produces strong UV absorption, which likely arises from simplified assumptions in the UV attenuation and absorption-line treatment, where a more sophisticated treatment of UV opacity and radiative transfer should be considered.

The similar break frequencies observed in the V-shaped SEDs of LRDs suggest a common physical mechanism underlying their unique SEDs. Setton et al. \cite{2024arXiv241103424Setton} proposed that LRDs exhibiting strong changes in spectral slope are highly unlikely to be explained by a two-component AGN model, which accounts for the break through a combination of scattered and attenuated components. In the self-regulated marginally unstable accretion disk model, the V-shaped SEDs arise from a composite spectrum contributed by the inner standard disk and the outer gravitationally unstable disk. Based on typical parameter distributions (e.g., $M_{\rm BH}\sim 10^{6.5-8.5}M_{\odot}$, $\dot{M}=0.01-0.5\dot{M}_{\rm Edd}$ and $R_{\rm out}=2-5R_{\rm c}$) used to model the LRD SEDs, we find that the break frequencies lie within a narrow range (e.g., $\sim 10^{15\pm0.1}\rm Hz$). Consequently, this model provides a natural explanation without the need to fine-tune numerous model parameters. For supermassive black holes with lower masses (e.g., $M_\mathrm{BH}\lesssim 10^{6.5}M_{\odot}$), the critical radius of the gravitationally unstable disk is larger in sub-Eddington states, resulting in a lower disk temperature ($\lesssim 1000$ K) and less pronounced infrared emission. If the accretion rates approach or exceed the Eddington limit, the radiation from the inner disk becomes stronger, while the radiation from the outer disk changes less due to the increase in $R_{\rm c}$, which also leads to a lower break frequency in the composite spectrum (see Method). Variability is a fundamental characteristic of AGNs and can serve as an indicator of fluctuations near the horizons of central supermassive black holes. In our model, the infrared-to-optical and optical-UV emissions originate from the outer disk and the inner disk, which suggests that the variability in infrared-to-optical and optical-UV wavebands may not be correlated on short timescales. Some studies have tentatively explored the variability in LRDs, but no significant variability has been reported \cite{2024arXiv240704777Kokubo, 2025ApJ...985..119Zhang, 2025ApJ...983L..26Tee}, despite several LRDs displaying potential variability \cite{2025ApJ...985..119Zhang}. Time-domain data from JWST on LRDs remain sparse, and multi-epoch observations for more LRDs across different wavebands, particularly in the ultraviolet band, can be used to test this model.

The break frequency of V-spectrum in our model is $\sim 10^{15}$ Hz, which is similar to the frequency for the Balmer break that was evidently detected in some LRDs \cite{2025ApJ...984..121Wang, 2024A&A...691A.310Kuruvanthodi, 2023ApJ...952..142Furtak}. Therefore, the absorption is sometimes assumed to be a reason for the Balmer break. The Balmer break may be indeed affected by the strong absorption in some LRDs (e.g., sources with Balmer break strength $>3$, as seen in UNCOVER 9497, CEERS 10444, PRIMER 33823 in our sample and the Cliff in \cite{2025A&A...701A.168deGraaff}), which is also supported by the observed Balmer absorption lines \cite{2025A&A...701A.168deGraaff, 2025ApJ...980L..29Akins, 2025arXiv250316595Rusakov, 2025arXiv250311752DEugenio}. However, many LRDs do not show a strong Balmer break, which does not support the V-shaped spectrum in all sources, and is caused by the absorption. Furthermore, the break frequency of V-shaped spectra in some LRDs clearly deviates from the Balmer limit (e.g., 1433-1045, JADES-13704 as discussed in Setton et al. \cite{2024arXiv241103424Setton}). This suggests that Balmer break may not be the primary physical mechanism for most V-shaped spectra in LRDs. We note that strong dust extinction is still required to reproduce the V-shaped spectra of LRDs in absorption scenarios (e.g., $A_V \sim 2-3$, \cite{2025ApJ...980L..27Inayoshi, 2025MNRAS.tmp.1770Ji}, but see \cite{2025arXiv250316596Naidu} for a different conclusion based on one LRD). However, the lack of hot and cold dust emission in most luminous LRDs does not support the dust-attenuated scenario \cite{2024ApJ...968...34Williams, 2025ApJ...991L..10Setton, 2025A&A...700A.231Xiao}. In our model, the infrared-to-optical spectral bump arises from emission in the outer gravitationally unstable disk. The strong Balmer break as found in several LRDs can be explained by the possible absorption from the dense gas surrounding the AGN disk, where the optical-UV emission may be suppressed by 2-3 times. The Balmer break is not very evident in most of the other LRDs, where the gas absorption should also not be important. Considering the possible disk wind scenario for the dense absorber, the LRDs with strong Balmer break and absorption features are possibly observed at large viewing angles along the motion of the winds, since the anisotropic disk wind is normally along the disk.  

\begin{table}
\centering
\caption{Best-fit parameters for the two typical LRDs.}
\label{tab:bestfit}

\setlength{\tabcolsep}{10pt}   
\scriptsize       

\renewcommand{\arraystretch}{2.0} 
\begin{tabular}{lcc}
\hline
\textbf{Parameter} & \textbf{UNCOVER 9497} & \textbf{PRIMER 33823} \\
\hline

$\log (\dot{M}/\dot{M}_{\rm Edd})$ 
    & $-1.26^{+0.01}_{-0.01}$ 
    & $-1.80^{+0.01}_{-0.01}$ \\

$R_{\rm out}/R_{\rm c}$ 
    & $3.39^{+0.03}_{-0.04}$
    & $2.85^{+0.01}_{-0.01}$ \\

$\log n_{\rm H}\,(\mathrm{cm^{-3}})$ 
    & $7.80^{+0.01}_{-0.01}$ 
    & $8.78^{+0.03}_{-0.03}$ \\

$\log N_{\rm H}\,(\mathrm{cm^{-2}})$ 
    & $24.97^{+0.02}_{-0.02}$ 
    & $23.85^{+0.01}_{-0.01}$ \\

\hline
\end{tabular}
\end{table}

The LRDs are typically X-ray weak sources, as evidenced by the nondetection of individual LRDs or the stacked sample in Chandra X-ray data \cite{2024ApJ...974L..26Yue, 2024ApJ...969L..18Ananna}. We estimate the X-ray emission for these LRDs based on the empirical correlation between optical and hard X-ray emissions (characterized by optical to X-ray spectral index \(\alpha_{\rm OX}\) and the X-ray spectral index $\Gamma$), where the optical emission is derived from our disk modeling of the optical to UV data. Our findings indicate that the X-ray intensity of most LRDs is below the detection limit of \textit{Chandra} (see Method for more details). We also estimate the line luminosities using the ionization spectrum from our disk model, based on photoionization calculations performed with CLOUDY simulations \cite{2023RMxAA..59..327Chatzikos}, where the locally optimally emitting clouds are considered with an assumed covering factor(CF, proportion of the sky obscured by the broad-line clouds). The CF value can be estimated by comparing the model result with the observed line luminosity (see Table 1). Excluding five sources with unreasonably large values of covering factor ($\rm CF>1$), we find that the average CF is approximately 0.3, which is roughly consistent with the typical value in normal AGNs \cite{2020MNRAS.494.5917Ferland}. The unreasonable covering factors in these five LRDs may be caused by not well-constrained absorptions or deviation of their spectra from the standard disk model (all of them have \(\dot{M}/\dot{M}_{\rm Edd} < 5\%\)). It should be noted that most LRDs are sub-Eddington sources and their intrinsic luminosities are not so high, which is the main difference for the dust-attenuated models. The total energy output based on multi-wavelength observations for these LRDs can provide a key discriminant between our model and alternative explanations.

The LRDs are common at high redshifts based on JWST observations. The evolution of the LRDs is still a puzzle, and it is unclear why no strong infrared-to-optical emission in other AGNs. The hot dust emission (\(\sim 500-2000 \, \text{K}\)) from the torus is notably absent in LRDs \cite{2025ApJ...991L..10Setton}, and this deficit may be attributed to low metallicities \cite{2024A&A...689A.220Tripodi, 2025ApJ...994...65Nakane}. Gravitational instability in the outer AGN disk provides a pathway for star formation. The embedded stars in AGN disks will accrete quickly in a dense environment and lead to fast metal enrichment after supernova explosion\cite{2023ApJ...944..159Fan}. The increase of metallicity with the evolution of AGNs will lead to more dust in the nuclear region of galaxies. The higher dust opacity will lead to a more inflated disk in the outer region, since the dust opacity sensitively depends on the metallicity \cite{2018MNRAS.474.1970Baskin}. If the optically-thick dust and gas sandwich the outer SG disk, the infrared-to-optical emission from the outer SG disk will be absorbed and re-radiated at longer wavelengths, where the inner radius of the torus is comparable to the size of the SG disk, as suggested by near-infrared reverberation mapping (e.g., \cite{1996ApJ...465L..87Nelson}). This may be the physical reason why the LRDs only appear in high-redshift dwarf galaxies, and the infrared-to-optical emission is suppressed in local AGNs and high-redshift QSOs with higher metallicities \cite{2018MNRAS.480..345Xu, 2020ApJ...898..105Onoue, 2021ApJ...910..115Sniegowska, 2022ApJ...925..121Wang, 2023ApJ...944..159Fan}. We note that seven local dwarf galaxies exhibiting V-shaped SEDs have been reported, which show sub-Eddington rates and are potentially local analogs to LRDs \cite{2025ApJ...980L..34Lin}. Further searches for possible low-redshift counterparts could provide valuable insight into the evolution of LRDs.

\section*{Methods}\label{method}
\bmhead{Photometric and Spectroscopic Data for LRDs}\label{sec:method1}

We analyze 28 confirmed LRDs with broad emission lines, primarily drawn from the sample of Kocevski et al. \cite{2025ApJ...986..126Kocevski}, supplemented by three additional objects (J0647–1045\cite{2024A&A...691A..52Killi}, COS 756434\cite{2025ApJ...991...37Akins}, and JADES 954\cite{2024A&A...691A.145Maiolino}). One source (PRIMER–UDS 119639) is excluded due to the lack of short-wavelength detections.

All photometric measurements are taken from publicly available JWST survey catalogs, which include the DAWN JWST Archive (DJA) grizli products\cite{brammer_2023_grizli, brammer_2023_msaexp} and the corresponding official survey data releases (e.g., UNCOVER DR3)\cite{2024ApJ...976..101Suess} and filter used listed in Supplementary Table \ref{suptable:supptable_1}. COS 756434 uses photometry compiled by \cite{2025ApJ...991...37Akins}. Spectroscopic data of NIRSpec/PRISM are selected from the public data of DJA products. Lensing magnification corrections are applied for UNCOVER sources using the provided magnification maps \cite{2023MNRAS.523.4568Furtak, 2025ApJ...982...51Price}.

In total, our dataset includes 28 LRDs with photometry and 23 with NIRSpec/PRISM spectra

\bmhead{The broken power-law fitting for the V-shaped SEDs of LRDs}\label{sec:method2}

We present the SEDs of our selected LRDs based on the photometric data shown in Supplementary Figure \ref{supfig:supp_figure_1}, where the V-shaped spectral features are evident in most of the sources. We initially fit the V-shaped SED in the rest frame with a simple broken-power-law model, where $\log {\nu L_\nu}=k_1 \cdot (\log \nu - \log \nu_{\rm b}) + \log {\nu L_\nu}_{\rm b}$ and  $\log {\nu L_\nu}=k_2 \cdot (\log \nu - \log \nu_{\rm b}) + \log {\nu L_\nu}_{\rm b}$ are adopted for the frequencies lower and higher than the break frequency, $\nu_{\rm b}$, respectively, by using curve\_fit \cite{2020NatMe..17..261Virtanen}. The fitting results are shown in Figure \ref{figure_1} for two typical LRDs. And the other are shown in Supplementary Figure \ref{supfig:supp_figure_1}. The fitting parameters are presented in Supplementary Table \ref{supfig:supp_figure_3}. For most LRDs, the fitting yields reasonable results, and the break frequencies of the V-shaped SEDs range from $10^{14.8}$ to $10^{15.1}$ Hz.

\bmhead{Model of the inner standard disk and the outer gravitationally unstable disk}\label{sec:method3}

The spectral energy distributions in this work are derived from the self-gravitating disk model as proposed by \cite{2003MNRAS.341..501Sirko}. The basic equations of this model are described as follows,

\begin{equation} \label{eq:sg_equation1}
\sigma T_{\mathrm{eff}}^4 = \frac{3}{8\pi}\dot{M}\Omega_{\rm K}^2\left[1-(\frac{R_{\rm in}}{R})^{1/2}\right]
\end{equation}
\begin{equation} \label{eq:sg_equation2}
T_{\rm c}^4 = \left( \frac{3}{8} \tau + \frac{1}{2} + \frac{1}{4\tau} \right) T_{\mathrm{eff}}^4,
\end{equation}
\begin{equation} \label{eq:sg_equation3}
\tau = \frac{\kappa \Sigma}{2},
\end{equation}
\begin{equation} \label{eq:sg_equation4}
\ c_s^2 \Sigma = \frac{\dot{M}\left[1-({R_{\rm in}}/{R})^{1/2}\right] \Omega_{\rm K}}{3 \pi \alpha},
\end{equation}
\begin{equation} \label{eq:sg_equation5}
p_{\mathrm{rad}} = \frac{\tau \sigma}{2c} T_{\mathrm{eff}}^4,
\end{equation}
\begin{equation} \label{eq:sg_equation6}
p_{\mathrm{gas}} = \frac{\rho k T_{\rm c}}{m},
\end{equation}
\begin{equation} \label{eq:sg_equation7}
\Sigma = 2 \rho h,
\end{equation}
\begin{equation} \label{eq:sg_equation8}
h = \frac{c_s}{\Omega_{\rm K}},
\end{equation}
\begin{equation} \label{eq:sg_equation9}
c_s^2 = \frac{p_{\mathrm{gas}} + p_{\mathrm{rad}}}{\rho},
\end{equation}
\begin{equation} \label{eq:sg_equation10}
\kappa = \kappa (\rho, T),
\end{equation}
where $\Omega_{\rm K}=\sqrt{GM_{\rm BH}/R^3}$ is Keplerian angular velocity, $R_{\rm in}$ is inner boundary of accretion disk, and $m=0.62 \, m_{\rm H}$ is the mean molecular mass. Given the black hole mass $M$, accretion rate $\dot{M_{\rm BH}}$, and viscosity parameter $\alpha$, these 10 equations can be solved, deriving 10 unknowns: effective temperature $T_\mathrm{eff}$, midplane temperature $T_{\rm c}$, optical depth $\tau$, surface density $\Sigma$, sound speed $c_\mathrm{s}$, radiation pressure $p_\mathrm{rad}$, gas pressure $p_\mathrm{gas}$, gas density $\rho$, disk half thickness $h$, and opacity $\kappa$ at any radius $R$.

In the outer parts of the disk, once Toomre parameter less than unity (i.e., $Q=c_\mathrm{s}\Omega_{\rm K}/\pi G \Sigma =\Omega_{\rm K}^2/2\pi G\rho<1$), the accretion disk is gravitationally unstable and prone to fragment into clumps or even stars \cite{1978AcA....28...91Paczynski, 2003MNRAS.339..937Goodman}, which may account for the supersolar metallicity of broad line regions \cite{1987Natur.329..810Shlosman, 2003MNRAS.339..937Goodman, 2023ApJ...954...84Wang, 2023ApJ...944..159Fan}. Sirko \& Goodman \cite{2003MNRAS.341..501Sirko} suggested that the outer disk could evolve toward a marginally stable state with $Q\sim1$. Chen et al. \cite{2023ApJ...948..120Chen} explored this issue and found that gravitationally unstable ($Q\sim1$) outer regions of AGN disks are possible based on 3D radiation hydrodynamic local shearing-box simulations. If this is the case, equation \ref{eq:sg_equation1} is no longer valid, the density of the outer disk is given by $\rho=\Omega_{\rm K}^2/2\pi G$, and $T_{\rm eff}$ should be derived from equation \ref{eq:sg_equation2}.

$R_\mathrm{in}=6\, R_\mathrm{g}$ and $\alpha=0.1$ are adopted as typical values throughout this work. For the dependency of opacity on density and temperature (equation \ref{eq:sg_equation10}), we use the opacity tables performed by \cite{1996ApJ...464..943Iglesias} for high temperature regimes ($> 10^{3.75} \, \mathrm{K}$). In the marginally stable region, the temperature of the disk is relatively low, and the opacity is dominated by dust grains. To account for this, we adopted the opacity tables for low temperature regimes ($< 10^{3.75} \, \mathrm{K}$) provided by \cite{2005ApJ...623..585Ferguson}, which takes into account the equation of state of the dust fully coupled to the gas.

With the above structure and temperature distribution of the accretion disk, the emitted spectrum is given by
\begin{equation}
    L_{\nu}=\pi\int_{R_{\rm in}}^{R_{\rm out}}\frac{2h\nu^3}{c^2}\frac{2\pi r}{{\rm exp}[h\nu/kT_{\rm eff}(R)] -1}dR, 
\end{equation}

The composite spectra and the temperature distribution from this accretion model for $M_{\rm BH}=10^{6-9}M_{\odot}$ and $\dot{M}=0.01-1\dot{M}_{\rm Edd}$ are shown in Supplementary Figure \ref{suptable:supptable_2}, where $R_{\rm out}=10R_{\rm c}$ is adopted. For $\dot{M}=0.01\dot{M}_{\rm Edd}$, the near infrared bump in the SED is less evident for $M_{\rm BH}\lesssim 10^{7}M_{\odot}$ (left panel) due to the lower disk temperature in outer marginally unstable disk (right panel). With an increase in accretion rate, the critical SMBH mass for an evident infrared bump becomes slightly lower (e.g., $M_{\rm BH}\lesssim 10^{6}M_{\odot}$ for $\dot{M}=0.3\dot{M}_{\rm Edd}$). For $\dot{M}\sim \dot{M}_{\rm Edd}$, the break frequencies also become slightly lower, as the optical/UV emissions become much stronger, while the near-infrared to optical emissions remain roughly unchanged due to the increase in $R_{\rm c}$ with increase of the accretion rate. Supplementary Figure \ref{suptable:supptable_2} shows that this model can naturally reproduce the V-shaped SEDs with break frequency $\sim 10^{15}$Hz except for some parameter space. This model predicts a black-body structure SED from near infrared to optical waveband, which is mainly contributed by the outer gravitationally unstable disk. The comparison of more LRDs with mid-infrared data can test this model and distinguish it from other models.

To account for the Balmer break as seen in some LRDs (e.g., UNCOVER 9497 and PRIMER 33823), we further consider the possible gas absorption as proposed by \cite{2025ApJ...980L..27Inayoshi}, who demonstrated that a Balmer break can naturally emerge if the AGN disk is embedded within extremely dense, neutral gas clumps ($n_\mathrm{H}\sim10^{9-11} \mathrm{cm}^{-3}$). Under this condition, collisional excitation populates hydrogen atoms into the $n=2$ state, which then efficiently absorbs the AGN continuum blueward of the Balmer limit, giving rise to both the Balmer break and associated absorption features in Balmer lines. We can derive the V-shaped SEDs with a possible Balmer discontinuity by considering the self-gravitating disk embedded in a dense gas.

\bmhead{The fitting of the V-shaped spectrum by the outer gravitationally unstable disk for all LRDs}\label{sec:method4}

For 20 LRDs with well-sampled spectroscopic data, we adopt the full four-parameter model that combines the self-gravitating accretion disk with absorption by dense gas. The intrinsic disk emission is generated by varying the accretion rate $\dot{M}$ and outer radius $R_{\mathrm{out}}$. These spectra are used as input to the photoionization code \textsc{Cloudy} (version C23.01; \cite{2023RMxAA..59..327Chatzikos}) interfaced via \textsc{pyCloudy} \cite{2013ascl.soft04020Morisset}, where we simulate transmission through a plane-parallel gas slab. The parameters used in the fitting are listed in Supplementary Table \ref{suptable:supptable_2}. In the fitting, we mask spectral regions below $\lambda<1216$ \AA, as well as regions near strong emission lines (e.g., $\rm H\alpha$, $\rm H\beta$, [O III]) and exclude photometric points with SNR $<1$. For the remaining 8 LRDs with incomplete spectral coverage, we fit their SEDs using a pure disk model without gas absorption, since their Balmer break feature cannot be well constrained. Fiiting results are shown in figure \ref{figure_5} The fitting results suggest that the spectrum of a self-gravitating disk alone can roughly reproduce the overall V-shaped continuum of most LRDs. For LRDs with strong Balmer breaks(e.g., $BBs\gtrsim2.5$), the inclusion of dense gas absorption is required to explain their discontinuity at the Balmer limit, where the UV continuum is suppressed by a factor of $\lesssim$ 2–3. For objects with weak Balmer breaks(e.g., $BBs<2.5$), gas absorption plays a less important role, and their UV spectra are dominated by the intrinsic disk emission. If the absorber is associated with disk winds, the observed diversity in Balmer break strengths may reflect orientation effects, where the polar direction will suffer less absorption. We note that the spectrum of J0647 1045 shows a strong Balmer jump feature, which is different from the simple V-shape power-law distribution that is also pointed out in \cite{2024arXiv241103424Setton}. Our model cannot fit the SED of this source very well.

\bmhead{Estimation of the X-ray strength in LRDs}\label{sec:method5}
The weak or non-detection of LRDs in X-ray wavebands suggests that their X-ray emission may be intrinsically weak or these LRDs are Compton-thick(e.g., \cite{2024ApJ...969L..18Ananna, 2024ApJ...974L..26Yue}). We estimate the strength of X-ray emission for these LRDs based on the disk emission as constrained from the the intrinsic model SED obtained from the fitting and the typical AGN SEDs from optical to X-ray wavebands, where the empirical correlation $\alpha_\mathrm{ox} = -0.14 \times \log L(2500\, \text{\AA})$ as constrained from type I AGNs\citep[e.g.,][]{2006AJ....131.2826Steffen} and typical unabsorbed X-ray power-law index of $\Gamma\sim1.7$ are adopted. We find that the predicted observation $\mathrm{2-10\ keV}$ X-ray flux range from $F_\mathrm{2-10\ keV}\sim10^{-17}-10^{-16}\ \mathrm{erg\ s^{-1}\ cm^{-2}}$ (see \textbf{Supplementary Table~\ref{suptable:supptable_1}}), which suggests that the X-ray emission of LRDs is intrinsically weak and nearly all LRDs in our sample are below the typical $Chandra$ detection limit(e.g., \cite{2009ApJS..184..158Elvis}). For the 8 sources where four-parameter fitting is not available, we use the results of the two-parameter fitting in our calculations by neglecting the absorption. We note that, even in objects with pronounced Balmer breaks, the optical–UV and X-ray continua are attenuated by only a factor of 2–3, which will not affect our main conclusion.

\bmhead{Estimation of the broad-line strength in LRDs}\label{sec:method6}
We also tentatively estimate the strength of the broad Balmer lines based on the photoionization model that is widely adopted to investigate the optical emission lines in AGNs, where the CLOUDY code version C23.01 is adopted with pyCloudy \cite{2023RMxAA..59..327Chatzikos, 2013ascl.soft04020Morisset}. We briefly describe the model here (see \cite{2024ApJ...965...84Wu} for more details). The locally optimally emitting cloud model was implemented to calculate emission line intensities, which assumes that substantial emission for any given spectral line originates exclusively within a narrow range of photoinization parameters (e.g., \cite{1995ApJ...455L.119Baldwin, 2003ARA&A..41..517Ferland}). The line flux from individual cloud surfaces is characterized by $F(r, n)$, where $r$ represents the distance from the central source and $n$ denotes the plasma number density. The total broad-line luminosities can be estimated from $L_\mathrm{line} \propto \iint r^2 F(r, n) f(r) g(n) \, \mathrm{d}n \mathrm{d}r $, where $f(r)$ and $g(n)$ describe the radial distribution and density distribution functions of clouds respectively. In our simulations, BLR clouds are modeled with solar metallicity and a hydrogen column density of $N_\mathrm{H} = 10^{23} \mathrm{cm}^{-2}$. The number density spans  $10^8-10^{14} \mathrm{cm}^{-3}$ with logarithmic grid steps of 0.1 dex for density and 0.2 dex for radial distance are adopted. As shown in the Method, we use the intrinsic SEDs as derived from our fitting model. To reproduce the observed line luminosities, we adjust the covering factor of the broad-line region clouds by assuming the model prediction to be the same as the observational line luminosity. In five sources, the derived CF exceeds unity, where these sources have very low Eddington ratios (a few percent or less), and their UV to hard X-ray ionizing spectrum may be underestimated if adopting the empirical correlations from the luminous AGNs. Excluding these five sources, we obtain an average CF of $\sim$0.3, which is consistent with typical values reported for type I AGNs \cite{2020MNRAS.494.5917Ferland}.

\backmatter

\noindent

\backmatter
\bmhead{Supplementary information}

Supplementary Information is available for this paper.

\bmhead{Acknowledgments}
CZ and QW are supported by the National Natural Science Foundation of China (NSFC, grants 12533005, 12233007), the China Manned Space Project (CMS-CSST-2025-A07), and the National Key Research and Development Program of China (No. 2023YFC2206702). XF is supported by the NSFC (12322307). LCH is supported by the NSFC (11991052, 12233001), the National Key R\&D Program of China (2022YFF0503401), and the China Manned Space Project (CMS-CSST-2021-A04, CMS-CSST-2021-A06). HZ acknowledges the National Science Foundation of China grant (No. 12303007). JW and XC are supported by the NSFC (12533005, 12233007, 12347103, and 12361131579), the science research grants from the China Manned Space Project (CMS-CSST-2025-A07) and the fundamental research fund for Chinese central universities (Zhejiang University). BL is supported by the NSFC (12133001).  J.M.W. thanks B-type Strategic Priority Program of the Chinese Academy of Sciences(XDB1160202), the NSFC (12333003), and the National Key R\&D Program of China(2021YFA1600404).

\bmhead{Author contributions} 
QW designed this project, CZ led the data analysis, CZ, XF, and JW took part in the modeling, LHC, XC, and JW took part in result interpretation, all authors contributed to manuscript discussion.

\bmhead{Data availability}

The reduced JWST data underlying this study are publicly available through the DAWN JWST Archive (DJA; \url{https://dawn-cph.github.io/dja/}). All raw JWST imaging and spectroscopic data used in this work are publicly available from the Mikulski Archive for Space Telescopes (MAST) under their respective survey program IDs. And were obtained from the following programs and fields: the JWST Advanced Deep Extragalactic Survey (JADES; Program IDs 1181 and 1210), the Public Release IMaging for Extragalactic Research survey (PRIMER; Program ID 4233), the COSMOS-Web Survey (JWST Cycle 2 Treasury Program; Program ID 6585), a General Observer program targeting the MACS J0647.7+7015 cluster (Program ID 1433), the Cosmic Evolution Early Release Science Survey (CEERS; Director’s Discretionary Early Release Science Program ID ERS-1345), and the Ultradeep NIRSpec and NIRCam Observations Before the Epoch of Reionization survey (UNCOVER; Program ID GO-2561).

\bmhead{Code availability}
All software packages used in the analysis are publicly available \texttt{pyCloudy}\cite{2013ascl.soft04020Morisset}, and \texttt{Cloudy}\cite{2023RMxAA..59..327Chatzikos}, and can be obtained from their respective public repositories. The code of the accretion disk is available by request from the corresponding author.

\bmhead{Competing interests} The authors declare no competing interests.

\bmhead{Corresponding author} Correspondence and requests for materials should be addressed to qwwu@hust.edu.cn (Qingwen Wu).
\clearpage

\section*{Supplementary Information}
\renewcommand{\figurename}{Supplementary Figure}
\renewcommand{\tablename}{Supplementary Table}
\setcounter{figure}{0}
\setcounter{table}{0}

\begin{table*}[ht]
\centering
\caption{\textbf{Summary of photometric filters used in this work.}}
\begin{tabular}{p{2.5cm} p{10cm}}
\toprule
\textbf{Survey} & \textbf{Photometric Filters} \\
\midrule

\textbf{UNCOVER} & 
\textit{JWST/NIRCam:} F070W, F090W, F115W, F140M, F150W, F162M, F182M, F200W, F210M, F250M, F277W, F300M, F335M, F356W, F360M, F410M, F430M, F444W, F460M, F480M; \\
& \textit{HST/ACS+WFC3:} F105W, F125W, F140W, F160W, F606W, F814W; \\
\midrule

\textbf{JADES} &
\textit{JWST/NIRCam:} F090W, F115W, F150W, F182M, F200W, F210M, F277W, F335M, F356W, F410W, F444W, F430M, F460M, F480M; \\
& \textit{HST/ACS+WFC3:} F435W, F606W, F775W, F814W, F850LP, F105W, F125W, F140W, F160W \\
\midrule

\textbf{CEERS} &
\textit{JWST/NIRCam:} F115W, F150W, F182M, F200W, F210M, F277W, F356W, F410M, F435W, F444W; \\
& \textit{HST/ACS+WFC3:} F105W, F125W, F140W, F160W, F606W, F814W;  \\
\midrule

\textbf{PRIMER} &
\textit{JWST/NIRCam:} F090W, F115W, F150W, F200W, F277W, F356W, F410W, F435W, F444W; \\
& \textit{JWST/MIRI:} F770W, F1800W; \\
& \textit{HST/ACS+WFC3:} F125W, F140W, F160W, F606W, F814W; \\
\midrule

\textbf{MACS0647} &
\textit{JWST/NIRCam:} F115W, F150W, F200W, F277W, F356W, F444W \\
\bottomrule
\label{suptable:supptable_1}
\end{tabular}
\end{table*}

\begin{table}[ht]
\centering
\caption{\textbf{Model parameters adopted in the CLOUDY simulations.}}
\label{tab:cloudy_params}
\large 
\begin{tabular}{p{0.4\columnwidth} p{0.55\columnwidth}}
\toprule
\textbf{Parameter} & \textbf{Value} \\
\midrule
$\log(Z/Z_{\odot})$ & $-1$ \\
$\log n_{\rm H}$ [cm$^{-3}$] & $7, 8, 9, 10, 11$ \\
$\log N_{\rm H}$ [cm$^{-2}$] & $22, 23, 24, 25$ \\
$v_{\rm turb}$ [km\,s$^{-1}$] & $120$ \\
Dust & none \\
\bottomrule
\label{suptable:supptable_2}
\end{tabular}
\end{table}

\begin{sidewaystable}
\centering
\setlength{\tabcolsep}{3pt}
\footnotesize
\caption{Source information and summary of fitting parameters, where the first 20 sources are fitted with four parameters ($\dot{m}$, $R_{\rm out}$, $n_{\rm H}$, and $N_{\rm H}$), whereas the last 8 sources are fitted with only two parameters ($\dot{m}$ and $R_{\rm out}$) without considering the absorption. Columns 1 to 4 represent the name, redshift, Balmer break strength (BBs), and BH mass, where the references for the BH mass are $a$) \cite{2023ApJ...959...39Harikane}; $b$) \cite{2023ApJ...954L...4Kocevski}; $c$) \cite{2025ApJ...986..126Kocevski}; $d$) \cite{2025ApJ...980L..29Akins}; $e$) \cite{2024A&A...691A.145Maiolino}; $f$) \cite{2024A&A...691A..52Killi}; $g$) \cite{2024ApJ...964...39Greene}. Note that the black hole masses of CEERS 69459 and the sources in the UNCOVER survey were derived from the intrinsic luminosity and line width of $\rm H\alpha$ emission in this work. From columns 5 to 9, we present the model fitting parameters of the break frequency, dimensionless accretion rate, the outer radius of the self-regulated disk, number density of absorber, and column density of absorber, respectively. In column 10, we list $\chi^2$ for the model fitting. In columns 11 and 12, we present the estimated covering factor of the broad emission line region and the predicted X-ray flux at the observational frame $\mathrm{2-10\ keV}$.}
\begin{tabular}{llcccccccccc}
\hline
Name & $z$ & BBs & $\log M_{\mathrm{BH}}$ & $\log \nu_{\mathrm{b}}$ & $\log\dot{m}$ & $R_{\mathrm{out}}$ & $\log n_{\rm H}$ & $\log N_{\rm H}$ & $\chi^2$ & CF & $F_\mathrm{2-10keV}$\\
& & & $(M_{\odot})$ & (Hz)  &     & $(R_{\mathrm{c}})$ & $(\rm cm^{-3})$ & $(\rm cm^{-2})$ & & & $(\mathrm{erg\ s^{-1}\ cm^{-2}})$\\
\hline
CEERS 82815 & 5.624 & 0.46 & 6.95($b$) & $14.89{\pm0.12}$ & $-0.88^{+0.01}_{-0.03}$ & $3.44^{+0.21}_{-0.23}$ & $7.73^{+0.34}_{-0.35}$ & $23.48^{+0.18}_{-0.24}$ & 2.24 & 0.69 & $1.36\times10^{-17}$\\
CEERS 5760 & 5.079 & 1.48 & 6.72($c$) & $15.02{\pm0.16}$ & $-0.48^{+0.01}_{-0.01}$ & $2.46^{+0.04}_{-0.03}$ & $8.75^{+0.45}_{-0.28}$ & $22.49^{+0.32}_{-0.25}$ &1.86 & 0.45 & $1.97\times10^{-17}$\\
CEERS 7902 & 6.986 & 2.40 & 8.24($c$) & $14.94{\pm0.06}$ & $-1.63^{+0.06}_{-0.01}$ & $3.10^{+0.15}_{-0.01}$ & $8.72^{+0.55}_{-0.58}$ & $22.78^{+0.36}_{-0.39}$ & 4.71 & -- & $4.97\times10^{-17}$\\
CEERS 10444 & 6.689 & 2.88 & 8.57($c$) & $15.02{\pm0.07}$ & $-1.85^{+0.01}_{-0.12}$ & $3.55^{+0.01}_{-0.44}$ & $8.61^{+0.07}_{-0.07}$ & $24.02^{+0.03}_{-0.02}$ & 5.58 & -- & $7.38\times10^{-17}$\\
CEERS 13135 & 4.955 & 3.90 & 7.39($c$) & $15.04{\pm0.06}$ & $-1.15^{+0.17}_{-0.04}$ & $2.57^{+0.15}_{-0.09}$ & $8.83^{+0.52}_{-0.71}$ & $23.39^{+0.15}_{-0.09}$ & 2.20 & 0.98 & $3.64\times10^{-17}$\\
CEERS 13318 & 5.280 & 2.09 & 8.34($c$) & $15.08{\pm0.06}$ & $-2.08^{+0.02}_{-0.08}$ & $2.19^{+0.04}_{-0.04}$ & $8.33^{+3.34}_{-0.31}$ & $23.31^{+2.83}_{-0.25}$ & 7.00 & 11.22 & $6.14\times10^{-17}$\\
CEERS 20496 & 6.786 & 1.60 & 6.45($c$) & $14.99{\pm0.15}$ & $0.68^{+0.05}_{-0.03}$ & $4.68^{+0.19}_{-0.11}$ & $8.31^{+0.31}_{-0.20}$ & $23.10^{+0.13}_{-0.78}$ & 1.52 & -- & $1.98\times10^{-17}$\\
CEERS 20777 & 5.286 & 0.84 & 6.84($c$) & $14.99{\pm0.09}$ & $-0.36^{+0.03}_{-0.05}$ & $2.90^{+0.07}_{-0.15}$ & $7.25^{+0.38}_{-0.19}$ & $23.41^{+0.14}_{-0.22}$ & 1.44 & 0.63 & $2.49\times10^{-17}$\\
PRIMER 29881 & 6.170 & 2.01 & 6.98($c$) & $14.91{\pm0.06}$ & $-0.72^{+0.06}_{-0.07}$ & $2.50^{+0.16}_{-0.10}$ & $8.04^{+0.65}_{-0.68}$ & $22.64^{+0.57}_{-0.45}$ & 1.24 & 0.69 & $16.5\times10^{-17}$\\
PRIMER 31092 & 5.675 & 0.85 & 7.10($c$) & $14.93{\pm0.07}$ & $-0.10^{+0.01}_{-0.04}$ & $4.14^{+0.05}_{-0.16}$ & $8.11^{+0.37}_{-0.17}$ & $23.23^{+0.09}_{-0.90}$ & 1.83 & 0.44 & $4.59\times10^{-17}$\\
PRIMER 33823 & 3.103 & 2.51 & 8.16($c$) & $15.07{\pm0.10}$ & $-1.80^{+0.01}_{-0.01}$ & $2.85^{+0.01}_{-0.01}$ & $8.78^{+0.03}_{-0.03}$ & $23.85^{+0.01}_{-0.01}$ & 15.88 & 7.41 & $1.96\times10^{-16}$\\
COS 756434 & 6.999 & 2.05 & 7.80($d$) & $14.94{\pm0.07}$ & $-1.34^{+0.01}_{-0.02}$ & $2.63^{+0.09}_{-0.02}$ & $9.00^{+0.49}_{-0.05}$ & $23.19^{+0.13}_{-0.46}$ & 5.87 & 0.30 & $3.07\times10^{-17}$\\
JADES 12068 & 5.919 & 1.28 & 7.50($e$) & $15.08{\pm0.07}$ & $-1.33^{+0.15}_{-0.02}$ & $1.89^{+0.11}_{-0.06}$ & $9.55^{+0.06}_{-1.29}$ & $22.25^{+1.15}_{-0.10}$ & 3.87 & 0.35 & $2.51\times10^{-17}$\\
JADES 954 & 6.760 & 1.36 & 7.90($e$) & $14.93{\pm0.05}$ & $-1.25^{+0.13}_{-0.07}$ & $2.98^{+0.52}_{-0.12}$ & $8.00^{+1.65}_{-0.88}$ & $23.17^{+0.46}_{-0.26}$ & 15.10 & 1.41 & $3.96\times10^{-17}$\\
J0647 1045 & 4.532 & 0.76 & 7.90($f$) & $15.02{\pm0.07}$ & $-1.36^{+0.01}_{-0.08}$ & $2.74^{+0.01}_{-0.25}$ & $7.80^{+1.20}_{-0.74}$ & $23.59^{+0.03}_{-0.02}$ & 35.04 & 0.54 & $7.24\times10^{-17}$\\
UNCOVER 3989 & 6.770 & 0.68 & 7.14($g$) & $14.91{\pm0.04}$ & $-0.46^{+0.03}_{-0.05}$ & $3.36^{+0.13}_{-0.21}$ & $8.32^{+0.22}_{-0.11}$ & $23.37^{+0.10}_{-0.12}$ & 1.62 & 0.37 & $2.40\times10^{-17}$\\
UNCOVER 4535 & 4.962 & 1.43 & 7.88($g$) & $15.01{\pm0.05}$ & $-1.71^{+0.01}_{-0.05}$ & $1.81^{+0.03}_{-0.08}$ & $7.92^{+2.24}_{-0.55}$ & $23.37^{+2.52}_{-0.23}$ & 2.15 & 1.45 & $4.81\times10^{-17}$\\
UNCOVER 9497 & 7.029 & 3.47 & 7.30($g$) & $14.98{\pm0.02}$ & $-1.26^{+0.01}_{-0.01}$ & $3.39^{+0.03}_{-0.04}$ & $7.80^{+0.01}_{-0.01}$ & $24.97^{+0.02}_{-0.02}$ & 6.19 & -- & $1.31\times10^{-17}$\\
UNCOVER 11119 & 5.835 & 1.21 & 7.65($g$) & $14.98{\pm0.06}$ & $-1.46^{+0.08}_{-0.01}$ & $2.49^{+0.07}_{-0.11}$ & $8.33^{+0.26}_{-0.43}$ & $23.27^{+0.08}_{-0.05}$ & 1.35 & 1.86 & $3.01\times10^{-17}$\\
UNCOVER 29255 & 8.504 & 2.02 & 7.81($g$) & $14.93{\pm0.03}$ & $-1.78^{+0.01}_{-0.01}$ & $3.86^{+0.09}_{-0.05}$ & $9.50^{+0.24}_{-0.37}$ & $22.13^{+0.16}_{-0.09}$ & 1.93 & -- & $1.30\times10^{-17}$\\
\hline
\hline
CEERS 69459 & 5.666 & -- & 6.66($a$) & $14.91{\pm0.03}$ & $-0.28^{+0.06}_{-0.08}$ & $2.38^{+0.12}_{-0.12}$ & -- & -- & 3.11 & 0.26 & $1.64\times10^{-17}$\\
CEERS 6126 & 5.288 & -- & 7.18($c$) & $15.00{\pm0.05}$ & $-0.45^{+1.34}_{-0.02}$ & $3.43^{+2.01}_{-0.12}$ & -- & -- & 11.48 & 0.49 & $ 4.31\times10^{-17}$\\
CEERS 14949 & 5.684 & -- & 6.95($c$) & $14.98{\pm0.05}$ & $-0.74^{+0.02}_{-0.01}$ & $2.74^{+0.03}_{-0.01}$ & -- & -- & 2.19 & 0.79 & $1.85\times10^{-17}$\\
PRIMER 32438 & 3.500 & -- & 6.98($c$) & $14.95{\pm0.08}$ & $-0.68^{+0.03}_{-0.02}$ & $2.27^{+0.06}_{-0.07}$ & -- & -- & 79.61 & -- & $5.59\times10^{-17}$\\
PRIMER 116251 & 5.365 & -- & 6.44($c$) & $14.93{\pm0.08}$ & $0.75^{+0.06}_{-0.08}$ & $4.67^{+0.34}_{-0.29}$ & -- & -- & 2.66 & 0.05 & $3.38\times10^{-17}$\\
UNCOVER 9358 & 6.354 & -- & 7.82($g$) & $14.94{\pm0.04}$ & $-1.86^{+0.01}_{-0.01}$ & $4.59^{+0.26}_{-0.38}$ & -- & -- & 103.52 & 1.91 & $2.20\times10^{-17}$\\
UNCOVER 9904 & 5.054 & -- & 7.83($g$) & $14.97{\pm0.06}$ & $-1.63^{+0.01}_{-0.01}$ & $2.08^{+0.01}_{-0.01}$ & -- & -- & 113.99 & 0.54 & $4.65\times10^{-17}$\\
UNCOVER 35436 & 6.741 & -- & 6.40($g$) & $14.92{\pm0.09}$ & $0.17^{+0.03}_{-0.03}$ & $2.75^{+0.08}_{-0.02}$ & -- & -- & 5.45 & 1.02 & $1.09\times10^{-17}$\\
\hline
\label{suptable:supptable_3}
\end{tabular}
\end{sidewaystable}

\clearpage

\begin{figure*}[ht]
    \centering
    \includegraphics[width=\textwidth]{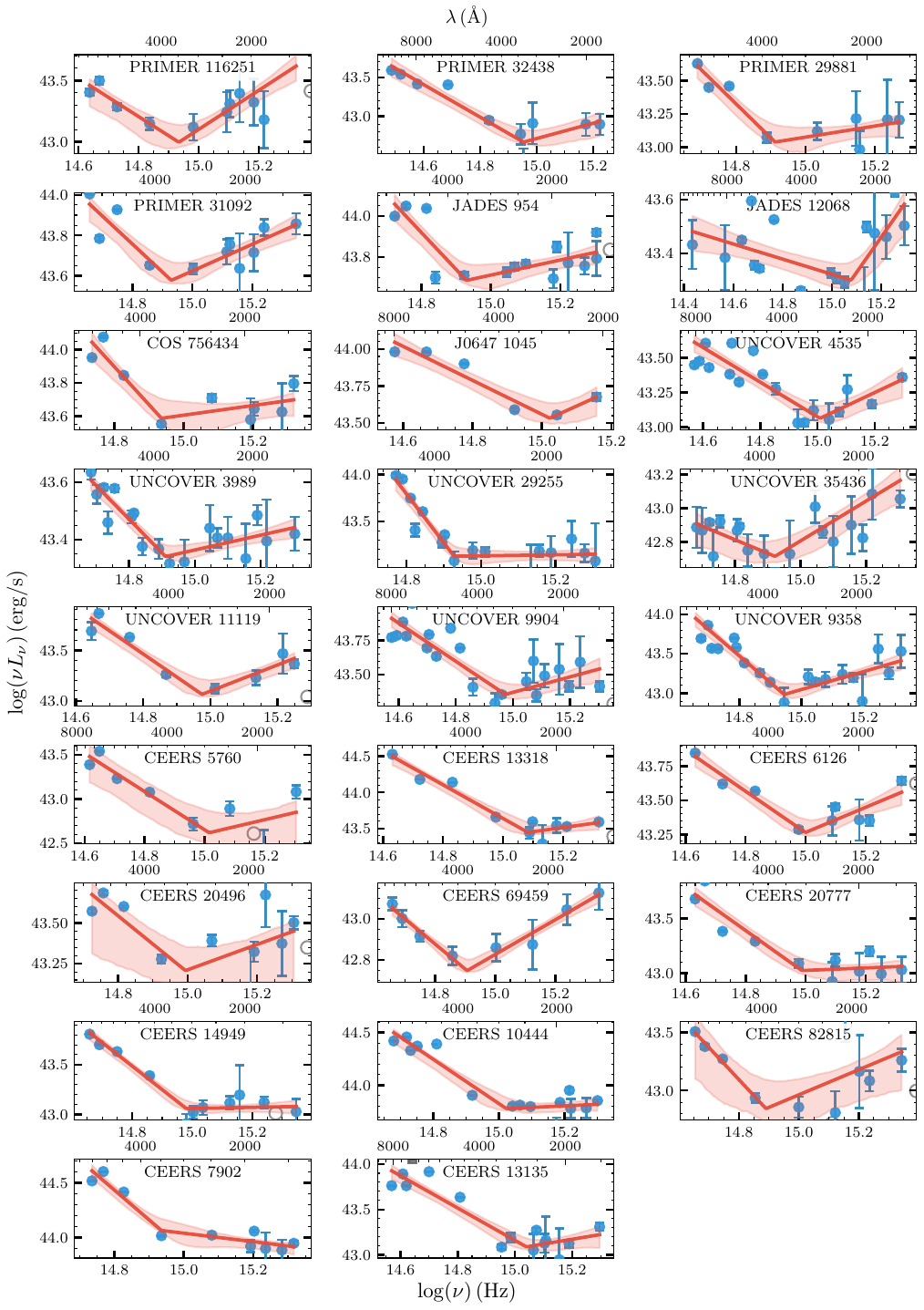}
    \caption{\textbf{Photometric SEDs and broken-power-law fits of LRDs.} The blue circles denote the photometric measurements used in the fitting with $1\sigma$ error bars. The V-shaped SEDs are fitted with a broken-power-law function (red line), where the shaded region represents $1\sigma$ uncertainty in the fitting. Points with low SNR or affected by emission line are shown in grey ponit.}
    \label{supfig:supp_figure_1}
\end{figure*}

\begin{figure*}[ht]
    \centering
    \includegraphics[width=\linewidth]{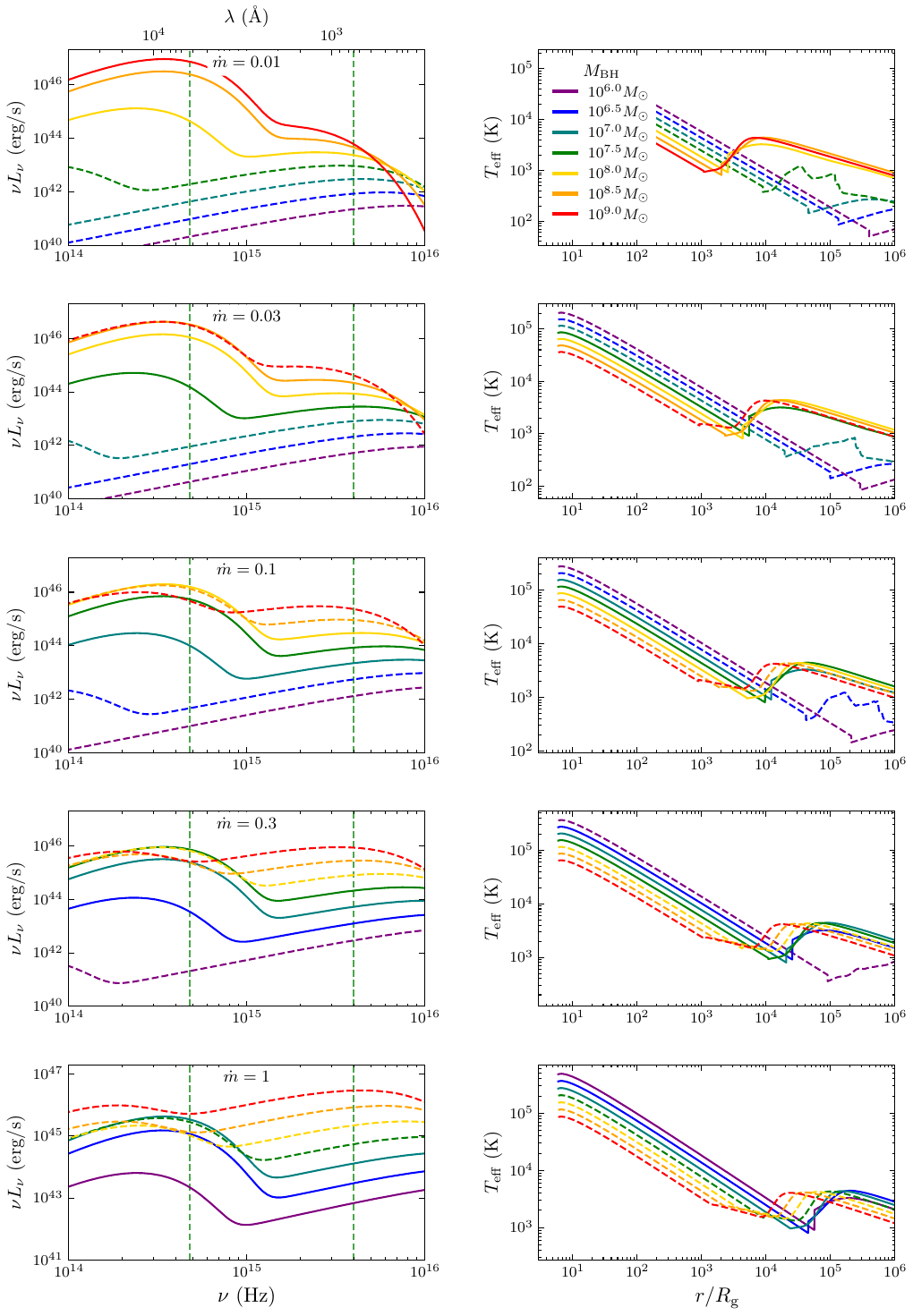}
    \caption{\textbf{Model SEDs and distribution of effective disk temperatures.} Left panels show model SEDs for different black hole masses at given accretion rates and $R_\mathrm{out}=10R_\mathrm{c}$ based on our standard inner disk and gravitationally unstable outer disk model; right panels show the effective disk temperature distributions. Vertical green lines indicate the NIRCam wavelength coverage at $z=7$.}
    \label{supfig:supp_figure_2}
\end{figure*}

\begin{figure*}[ht]
    \centering
    \begin{subfigure}[b]{0.6\textwidth}
        \centering
        \includegraphics[width=\textwidth]{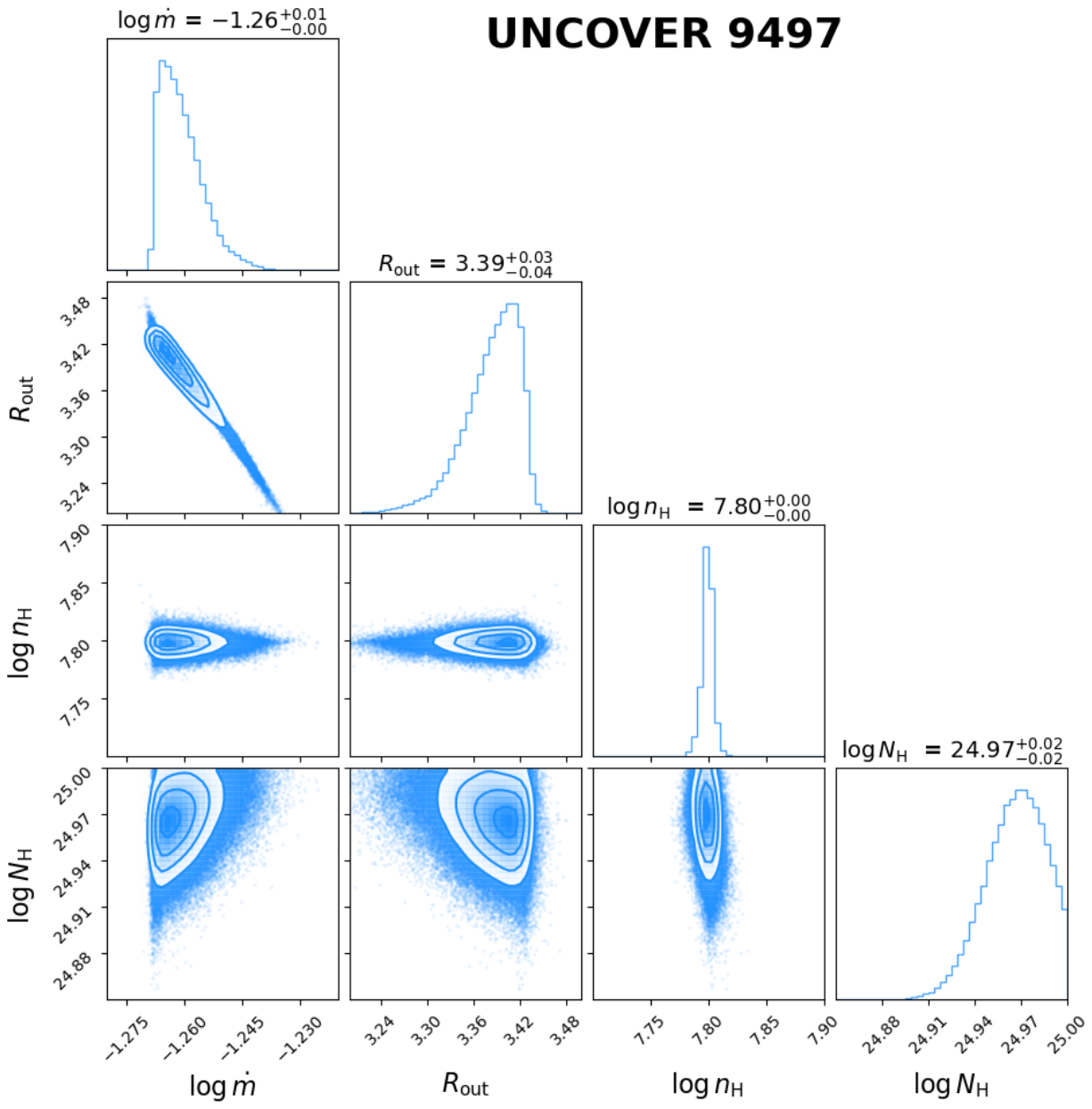}
    \end{subfigure}
    \hfill
    \begin{subfigure}[b]{0.6\textwidth}
        \centering
        \includegraphics[width=\textwidth]{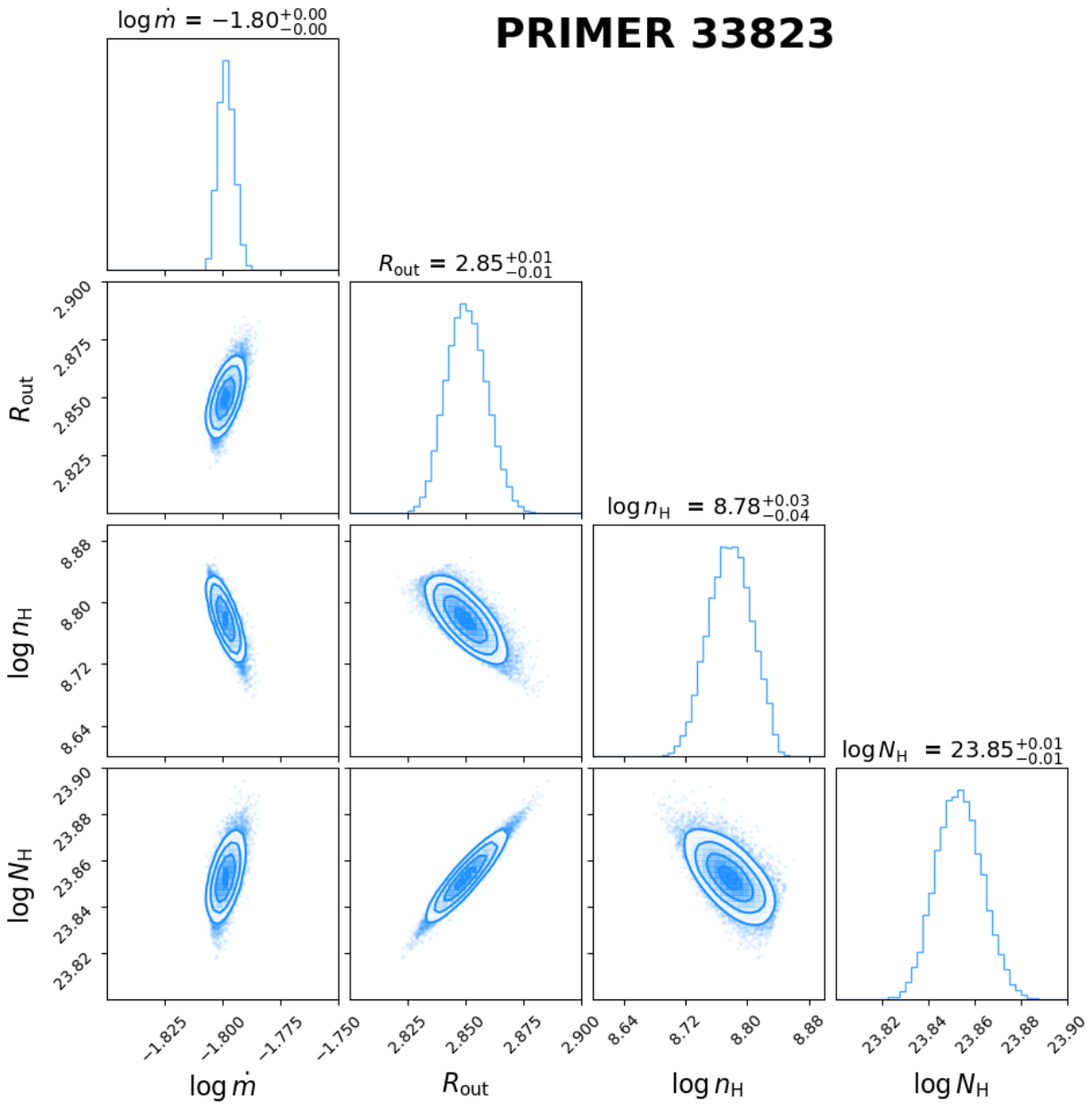}
    \end{subfigure}
    \caption{\textbf{Posterior distributions of accretion-disk parameters for two typical LRDs.} The posterior probability distributions for dimensionless accretion rate, outer disk radius, hydrogen density, and column density are represented for PRIMER 33823 and UNCOVER 9497 respectively.}
    \label{supfig:supp_figure_3}
\end{figure*}

\clearpage

\bibliography{sn-bibliography}

\end{document}